\documentclass[aps,pra,reprint,superscriptaddress]{revtex4-2}

\usepackage{amsmath,amsfonts,amssymb}
\usepackage{xcolor,graphicx}
\usepackage{braket}
\usepackage[pdftex]{hyperref}
\hypersetup{
	colorlinks = true,
	linkcolor = blue,
	anchorcolor = blue,
	citecolor = red,
	filecolor = blue,
	urlcolor = blue}

\begin{document}

\title{Asymmetric {C}auchy-{R}iemann Beams}

\author{N. Korneev}
\affiliation{Instituto Nacional de Astrofísica Óptica y Electrónica, Calle Luis Enrique Erro No. 1\\ Santa María Tonantzintla, Puebla, 72840, Mexico}

\author{I. Ramos-Prieto}
\email[e-mail: ]{iran@inaoep.mx}
\affiliation{Instituto Nacional de Astrofísica Óptica y Electrónica, Calle Luis Enrique Erro No. 1\\ Santa María Tonantzintla, Puebla, 72840, Mexico}

\author{I. Julián-Macías}
\affiliation{Instituto Nacional de Astrofísica Óptica y Electrónica, Calle Luis Enrique Erro No. 1\\ Santa María Tonantzintla, Puebla, 72840, Mexico}

\author{U. Ru\'iz}
\affiliation{Instituto Nacional de Astrofísica Óptica y Electrónica, Calle Luis Enrique Erro No. 1\\ Santa María Tonantzintla, Puebla, 72840, Mexico}

\author{F. Soto-Eguibar}
\affiliation{Instituto Nacional de Astrofísica Óptica y Electrónica, Calle Luis Enrique Erro No. 1\\ Santa María Tonantzintla, Puebla, 72840, Mexico}

\author{D. {S\'anchez}-{de-la-Llave}}
\affiliation{Instituto Nacional de Astrofísica Óptica y Electrónica, Calle Luis Enrique Erro No. 1\\ Santa María Tonantzintla, Puebla, 72840, Mexico}

\author{H.M. Moya-Cessa}
\affiliation{Instituto Nacional de Astrofísica Óptica y Electrónica, Calle Luis Enrique Erro No. 1\\ Santa María Tonantzintla, Puebla, 72840, Mexico}

\date{\today}

\begin{abstract}
    We investigate, theoretically and experimentally, the evolution of a paraxial beam propagating in free space  when its initial transverse structure is characterized by an asymmetric Gaussian modulation combined with an entire function. Utilizing a quantum optics operator approach, our study specifically examines the effects of parameter variations within the Gaussian modulation on two entire functions: the Bessel function and the Airy function. Through this investigation, we aim to elucidate how these parameter variations influence the beam's propagation dynamics and the role played by the asymmetry of the Gaussian modulation in the propagation of such paraxial beams. Additionally, we provide a comprehensive method for computing the propagated field under these conditions.
\end{abstract}
\maketitle

It was recently demonstrated that when a beam satisfies the Cauchy-Riemann equations at $z=0$ and is symmetrically modulated by a Gaussian function, its paraxial propagation in free space along the propagation distance $z$ can be straightforwardly derived using an operator approach from quantum optics~\cite{Moya_2024}. Furthermore, in a quadratic gradient-index medium, the same approach has yielded a simple expression for the two-dimensional fractional Fourier transform of an entire function~\cite{Ramos_2024}. Given that both the free particle problem and free space propagation are governed by the same equation, classical optics has explored diverse initial conditions, replicating a range of beams known as paraxial beams~\cite{Abramochkin_1993, Roux_2003, Kiselev_2004, Kiselev_2007, Bandres_2007, Kotlyar_2007}. For example, the non-trivial nature of the evolution of a free particle is widely acknowledged. Specifically, Berry and Balazs demonstrated that a particle with an initial Airy profile undergoes bending throughout its evolution, even in the absence of any acting potential~\cite{Berry_1979}. This phenomenon can be attributed, in principle, to the Bohm potential, which leads to such "anomalous" evolution~\cite{Bohm_1952, Hojman_2021, Hojman_2022, Rozenman_2023, Silva_2023}.

However, in the case of asymmetric Gaussian modulation, i.e., $g_x \neq g_y \rightarrow e^{-(g_x x^2 + g_y y^2)}$, the direct application of the operator approach described in~\cite{Stoler_1980,Moya_2024, Ramos_2024,Korneev_2024} may not be applied as the asymmetry introduces significant mathematical complexity due to the different scaling in the $x$ and $y$ coordinates that complicates the spectral analysis and eigenfunction structure. Consequently, conventional symmetric operator methods must be adapted. This highlights the need for a different prior approach to effectively handle the complexities introduced by asymmetric Gaussian modulation. This is one of the main motivations of this letter, where we explore two representative examples: the Bessel function and the Airy function.

\textit{Quasi-{C}auchy-{R}iemann beams.} In the framework of the paraxial approximation, the evolution equation for light in free space is represented by:
\begin{equation}\label{0020}
\nabla_\perp^2 E(x,y,z) + i2k\frac{\partial E(x,y,z)}{\partial z} = 0,
\end{equation}
with$\nabla_\perp^2 = \frac{\partial^2}{\partial x^2} + \frac{\partial^2}{\partial y^2}$, and its solution can be expressed as:
\begin{equation}\label{propagation}
E(x,y,z) = e^{-\frac{i}{2k}z(\hat{p}_x^2+\hat{p}_y^2)} E(x,y,0),
\end{equation}
where $k = \frac{2\pi}{\lambda}$ denotes the wave number, and $E(x,y,0)$ represents the field amplitude at the plane $z = 0$. It is worth mentioning that we have adopted the mathematical framework of quantum optics, introducing the operators: $\hat{p}_x = -i\frac{\partial}{\partial x}$ and $\hat{p}_y = -i\frac{\partial}{\partial y}$, which satisfy the following commutation relations: $[x, \hat{p}_x] = [y, \hat{p}_y] = i$ and $[x, y] = [x, \hat{p}_y] = [y, \hat{p}_x] = [\hat{p}_x, \hat{p}_y] = 0$. Consequently, if at $z = 0$ the initial field is modulated by a Gaussian function, such that
\begin{equation}\label{initialc}
E(x,y,0) = e^{-(g_xx^2+g_yy^2)}\mathcal{E}(x+iy),
\end{equation}
where $g_j$ generally takes values in the complex number domain (with $j = x, y$), and $\mathcal{E}(x+iy)$ is an entire function, i.e., $\nabla_\perp^2 \mathcal{E}(x+iy) = 0$. From Eqs.(\ref{propagation}) and (\ref{initialc}), and given that the set of operators $\hat{p}_q^2$, $q^2$, and $q\hat{p}+\hat{p}q$ (with $q=x,y$) is closed under commutation, it can be readily shown that the field at a distance $z$ can be expressed as~\cite{Moya_2024}:
\begin{equation}\label{Psi_t}
E(x,y,z) = \prod_{q = x,y} e^{\alpha_q(z)q^2} e^{\beta_q(z)(q\hat{p}_q + \hat{p}_q q)} e^{\gamma_q(z)\hat{p}_q^2} \mathcal{E}(x+iy),
\end{equation}
where
\begin{equation}
\begin{split}
\alpha_q(z) & = -\frac{g_q}{w_q(z)}, \\
\beta_q(z) & = -\frac{\pi}{4} - \frac{i}{2} \ln[iw_q(z)], \\
\gamma_q(z) & = -\frac{g_qZ^2}{w_q(z)},
\end{split}
\end{equation}
with $w_q(z) = 1 + \frac{i2g_qz}{k}$ (with $q = x, y$).

For instance, when the Gaussian modulation is symmetric, i.e., $g_x = g_y = g$, the property that $\mathcal{E}(x + iy)$ is an entire function, meaning $\nabla_\perp^2 \mathcal{E}(x + iy) = 0$, can be directly applied to Eq.(\ref{Psi_t}), leading to the result~\cite{Moya_2024}
\begin{equation}
E\left(x, y, z\right) = \frac{e^{-\frac{g\left(x^2 + y^2\right)}{w(z)}}}{w(z)} \mathcal{E}\left(\frac{x + iy}{w(z)}\right).
\end{equation}
The term $\gamma(z) = \gamma_x(z) = \gamma_y(z)$ is omitted from the final expression because $e^{\gamma(z)(\hat{p}_x^2 + \hat{p}_y^2)} \mathcal{E}(x + iy) = \mathcal{E}(x + iy)$, which directly follows from the condition $\nabla_\perp^2 \mathcal{E}(x + iy) = 0$. However, this relationship is not applicable when $g_x \neq g_y$. To address this issue, it is necessary to rewrite the last term in Eq.~(\ref{Psi_t}) as follows:
\begin{equation}
e^{\gamma_x(z)\hat{p}_x^2} e^{\gamma_y(z)\hat{p}_y^2} = e^{[\gamma_x(z) - \gamma_y(z)]\hat{p}_x^2} e^{\gamma_y(z)(\hat{p}_x^2 + \hat{p}_y^2)}.
\end{equation}
As a result, we obtain:
\begin{equation}\label{final}
    \begin{split}
    E\left(x,y,z\right) &= \left(\prod_{q= x,y} e^{\alpha_q(z)q^2} e^{\beta_q(z)(q\hat{p}_q + \hat{p}_q q)}\right)\\&\times e^{\Delta(z) \hat{p}_x^2} \mathcal{E}(x + iy),
    \end{split}
\end{equation}
with $\Delta(z) = \gamma_x(z)-\gamma_y(z)$, and where we have used that $e^{\gamma_y(z)(\hat{p}_x^2+\hat{p}_y^2)}\mathcal{E}(x+iy)=\mathcal{E}(x+iy)$. This parameter, $\Delta(z)$, plays a pivotal role in determining the nature of the beam evolution over the distance propagation, particularly in scenarios where asymmetry between the two spatial dimensions is present. 

Specifically, when $\Delta(z) \ne 0$, it becomes evident that employing an entire function simplifies the two-dimensional propagation to a one-dimensional propagation. This simplification occurs because the propagation term is reduced from the exponential of the Laplacian in two variables to a single variable, $e^{\Delta(z) \hat{p}_x^2}$. This simplification can be more easily understood by expressing $\mathcal{E}(x+iy)=e^{-y \hat{p}_x}\mathcal{E}(x)$. Consequently, we can rewrite Eq.~(\ref{final}) as
\begin{equation}\label{eq:0080}
    \begin{split}
E\left(x,y,z\right)&=\left(\prod_{q= x,y}e^{\alpha_q(z)q^2}e^{\beta_q(z)(q\hat{p}_q+\hat{p}_qq)}\right)\\&\times e^{\Delta(z)\hat{p}_x^2} e^{-y\hat{p}_x}\mathcal{E}(x).
    \end{split}
\end{equation}
Expressing the function $\mathcal{E}(x)$ in terms of its Fourier transform, i.e.,
\begin{equation}
\mathcal{E}(x)=\int_{-\infty}^{\infty}\tilde{\mathcal{E}}(u)e^{iux}du,
\end{equation}
it is possible to obtain, after employing Hadamard's lemma and the Baker-Campbell-Hausdorff formula~\cite{RossmannW,Hall_2013}, that
\begin{equation}\label{GeneralCase}
    \begin{split}
E\left(x,y,z\right)&=\frac{e^{-\left(\frac{g_xx^2}{w_x(z)}+\frac{g_yy^2}{w_y(z)}\right)}}{\sqrt{w_x(z)w_y(z)}}\\
&\times\int\limits_{-\infty}^{\infty}\tilde{\mathcal{E}}(u)e^{iu\left(\frac{x}{w_x(z)}+i\frac{y}{w_y(z)}\right)}e^{-\Delta(z)u^2}du.
    \end{split}
\end{equation}
The above equation demonstrates that starting with any function $\mathcal{E}(x+iy)$ as an initial condition, we can compute its Fourier transform $\tilde{\mathcal{E}}(u)$ and then use it in the equation to determine the field at distance $z$. This method represents a valuable approach to obtain the evolution of any function $\mathcal{E}(x+iy)$ that satisfies the Cauchy-Riemann equations. Furthermore, it optimizes the numerical computation of this evolution process by first computing the fast Fourier transform, followed by integrating this transform using Eq.~(\ref{GeneralCase}). Now, we proceed to demonstrate several analytic solutions for Eq.~(\ref{final}). Specifically, we will explore solutions involving Bessel and Airy functions, which are widely studied special functions with significant applications across various fields of physics and engineering~\cite{mi11110997, Efremidis19}.

\begin{figure}
		\centering\includegraphics[width = \linewidth]{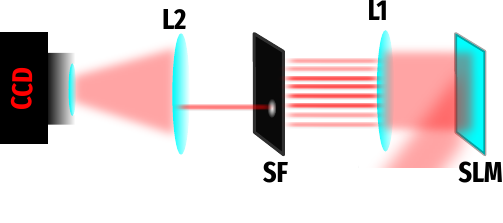}
		\caption{To synthesize the desired optical field, the experimental setup utilizes a linearly polarized Helium-Neon (He-Ne) laser with a wavelength of $\lambda = 633~\text{nm}$. The laser beam is directed towards a Spatial Light Modulator (SLM), which encodes the synthetic phase hologram. A 4f-optical system is employed, comprising two lenses that perform a Fourier transform of the SLM's output at the focal plane of the second lens. At this focal plane, a binary Spatial Filter (SF) is placed to isolate a specific diffracted order from the hologram. This selected order carries the desired complex field spatial distribution, facilitating the precise synthesis of the optical field through optical Fourier transformation, as detailed in Arrizón et al.~\cite{Arrizon07}. This setup ensures that only the intended field component is transmitted, effectively eliminating unwanted diffraction orders and enhancing the fidelity of the synthesized optical field.}
		\label{Fig_0}
\end{figure}

\textit{Bessel function.} We analyze the propagation of a Bessel function of the first kind $J_n(\eta x)$, with $\eta$ representing a scaling parameter. We use the following integral representation of the Bessel function,
\begin{equation}\label{Bessel}
J_n(\eta x)=\frac{1}{2\pi}\int\limits_{-\pi}^{\pi}e^{inu}e^{-i\eta  x\sin u}du.
\end{equation}
From Eq.~(\ref{eq:0080}), by commuting the last two operators and considering $\mathcal{E}(x) = J_n(\eta x)$, along with the integral representation mentioned earlier, we obtain:
\begin{equation}
    \begin{split}
    E\left(x,y,z\right)&=\left(\prod_{q= x,y}
    e^{\alpha_q(z)q^2}e^{\beta_q(z) \left(q \hat{p}_q+\hat{p}_q q\right)}\right)\\ 
    &\times \frac{1}{2\pi}e^{-y\hat{p}_x}e^{\Delta(z)\hat{p}_x^2}\int\limits_{-\pi}^{\pi}e^{inu}e^{-i\eta  x\sin u}du.
    \end{split}    
\end{equation}
This leads us to introduce the operator $e^{\Delta(z)\hat{p}_x^2}$ into the integral. Leveraging the fact that the exponential function acts as an eigenfunction of momentum, we establish that
\begin{equation}
e^{\Delta(z)\hat{p}_x^2}e^{-i\eta  x\sin u}
=e^{\frac{\Delta(z)\eta^2}{2}}e^{-\frac{\Delta(z)\eta^2}{2}\cos(2u)}e^{-i\eta x\sin u}.
\end{equation}
Writing $\cos(2u)=(e^{2iu} +e^{-2iu})/2 $, expanding the function $e^{-\frac{\Delta(z)\eta^2}{2}\cos(2u)}$ in its Taylor series , and using the binomial theorem, we can demonstrate that
\begin{equation}
    \begin{split}
e^{-\frac{\Delta(z)\eta^2}{2}\cos(2u)}&=\sum_{k=0}^{\infty}\frac{(-\Delta \eta^2)^k}{4^kk!}\\&\times\sum_{m=0}^k \frac{k!}{m!(k-m)!}e^{2iu(k-2m)}.
    \end{split}
\end{equation}
The summation over $m$ above can be extended to infinity, because all the added terms are null as they have a Gamma function of negative integers in the denominator; then, using the integral representation of the Bessel functions Eq.~(\ref{Bessel}), it can be derived that
\begin{equation} 
    \begin{split}
e^{\Delta(z)\hat{p}_x^2}J_n\left(\eta x \right) &= e^{\frac{\Delta(z)\eta^2}{2}}\\ 
&\times\sum_{l=-\infty}^{\infty}J_{n+2l}(\eta x)\sum_{m=0}^{\infty}
\frac{(-\Delta \eta^2)^{l+2m}}{4^{l+2m}(l+m)!m!},
    \end{split}
\end{equation}
where initially we set $j = k-m$, and subsequently, we set $l=j-m$. The second sum in the above expression corresponds to a Bessel function of order $l$, 
\begin{equation}
J_l(x) = \sum_{m=0}^\infty \frac{(-1)^m}{m!\, (m+l)!} {\left(\frac{x}{2}\right)}^{2m + l},
\end{equation}
and therefore, we obtain
\begin{equation} 
e^{\Delta(z)\hat{p}_x^2}J_n\left(\eta x \right) =e^{\frac{\Delta(z)\eta^2}{2}} 
\sum_{l=-\infty}^{\infty}i^lJ_{n+2l}(\eta x)J_l\left(i\frac{\Delta\eta^2}{2}\right).
\end{equation}
The above sum is recognized as a generalized Bessel function \cite{Dattoli_1990}
\begin{equation}
\mathcal{J}_n(x,y;s)=\sum_{l=-\infty}^{\infty}(-s)^lJ_{n+2l}( x)J_l\left(y\right).
\end{equation}
Subsequently, by applying the remaining operators as outlined in~\cite{Moya_2024}, we ultimately derive
\begin{equation}\label{PropagationJ}
\begin{split}
E\left(x,y,z\right)&=\frac{e^{-\left(\frac{g_xx^2}{w_x(Z)}+\frac{g_yy^2}{w_y(z)}\right) }}{\sqrt{w_x(z)w_y(z)}}\\&\times e^{-{\eta^3\Delta(z)}\left[\frac{x}{w_x(z)}+i\frac{y}{w_y(z)}+\frac{2\eta^3\Delta^2(z)}{3}\right]} 
\\&\times 
\mathcal{J}_n\left[\eta\left(\frac{x}{w_x(z)}+i\frac{y}{w_y(z)}\right),\,i\frac{\Delta\eta^2}{2};\,-i\right].
\end{split}
\end{equation}
\begin{figure}
\centering\includegraphics[width = 8cm]{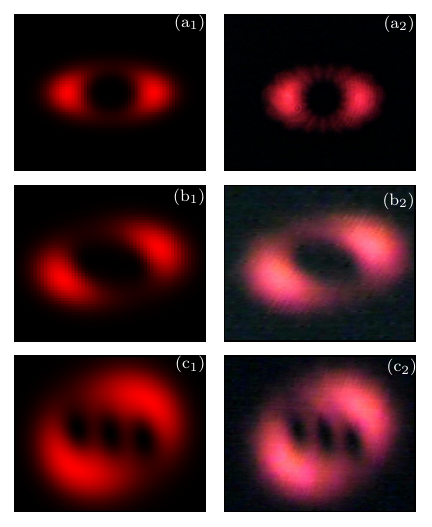}
		\caption{The intensity distribution of the field characterized by $J_n[\eta(x+iy)]$ with $n=3$ is shown at three different transverse planes: $(\text{a}_1)$ at $z=0.0\,\text{m}$, $(\text{b}_1)$ at $z=0.20\,\text{m}$, and $(\text{c}_1)$ at $z=0.90\,\text{m}$, as per Eq.~(\ref{PropagationJ}). Experimental observations for these planes are displayed in $(\text{a}_2)$, $(\text{b}_2)$, and $(\text{c}_2)$. The experimental parameters include $\eta = 3.57 \times 10^2\,\text{m}^{-2}$, $\lambda = 633\,\text{nm}$, $g_x = 0.5 \times 10^{7}\,\text{m}^{-2}$, and $g_y = 1 \times 10^{7}\,\text{m}^{-2}$, where $g_y/g_x=2.0$. The observation window is 8 millimeters wide. The experimental setup is illustrated in Fig.~\ref{Fig_0}.
        }
		\label{Fig_1}
\end{figure}
In Fig.~\ref{Fig_1}, we illustrate the propagation of a Bessel function, $\mathcal{E}(x+iy)=J_n[\eta(x+iy)]= e^{-y\hat{p}_x}J_n(\eta x)$ with $n=3$, across three distinct transverse observation planes: $(\text{a}_1)$-$(\text{a}_2)$, $(\text{b}_1)$-$(\text{b}_2)$, and $(\text{c}_1)$-$(\text{c}_2)$. Specifically, the first column in Fig.~\ref{Fig_1} represents the intensity distribution, $|E(x,y,z)|^2$, as derived from Eq.~(\ref{PropagationJ}), at three different distances along the $z$-axis: $(\text{a}_1)$ at $z=0\,\text{m}$, $(\text{b}_1)$ at $z=0.20\,\text{m}$, and $(\text{c}_1)$ at $z=0.90\,\text{m}$. In contrast, the second column in Fig.~\ref{Fig_1} depicts the intensity evolution at the same distances and under the same initial conditions, but for the experimentally synthesized optical field as detailed in Fig.~\ref{Fig_0}. This comparative analysis highlights that the number of intensity zeros is directly influenced by the parameter $n$ in the Bessel function $J_n[\eta(x+iy)]$. Moreover, the asymmetry in the Gaussian modulation, characterized by a ratio of $g_y/g_x=2.0$, results in a spatial separation of these intensity zeros within the propagated field. Such spatial distribution is vital for applications such as particle trapping and manipulation, where precise control over particle positioning and separation is crucial~\cite{Ashkin_1987}. The close correspondence between theoretical predictions and experimental results not only validates the theoretical model but also underscores its practical significance, thereby confirming the efficacy of the optical field synthesis technique~\cite{Arrizon07}. This validation is especially relevant for advancing optical technologies, including quantitative phase imaging~\cite{Oh_2023}, which utilizes complex optical fields of the form $\mathcal{E}(x+iy)$.
\begin{figure}
\centering\includegraphics[width = 8cm]{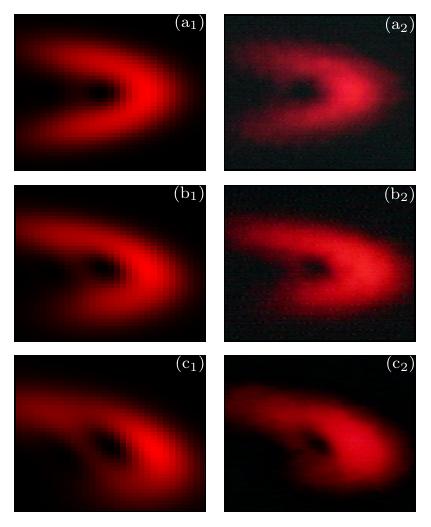}
		\caption{The intensity distribution of the field described by $\mathrm{Ai}[\eta(x+iy)]$ is presented at three transverse planes: $(\text{a}_1)$ at $z=0.0\,\text{m}$, $(\text{b}_1)$ at $z=0.20\,\text{m}$, and $(\text{c}_1)$ at $z=0.90\,\text{m}$, as specified by Eq.(\ref{Airy}). Corresponding experimental data for these planes are shown in $(\text{a}_2)$, $(\text{b}_2)$, and $(\text{c}_2)$. The experimental parameters are $\eta = 1.0 \times 10^3\,\text{m}^{-2}$, $\lambda = 633\,\text{nm}$, $g_x = 0.10 \times 10^{6}\,\text{m}^{-2}$, and $g_y = 0.50 \times 10^{6}\,\text{m}^{-2}$, with a ratio of $g_y/g_x=5.0$. The observation window width is 8 millimeters. The experimental setup is depicted in Fig.\ref{Fig_0}.
        }
		\label{Fig_2}
\end{figure}

\textit{Airy function.} As a further illustrative example, we consider the scenario where $\mathcal{E}(x+iy)$ is represented by an Airy function,
\begin{equation}
\mathcal{E}(x+iy)=\mathrm{Ai}[\eta(x+iy)]=e^{-y\hat{p}_x}\mathrm{Ai}(\eta x),
\end{equation}
where $\eta$ denotes a scaling parameter. Within the framework of wave-packet dynamics, Berry and Balazs~\cite{Berry_1979} demonstrated that the free evolution of such a wave-packet leads to the expression
\begin{equation}
e^{-i\frac{z}{2}\hat{p}_x^2} \mathrm{Ai}(\eta x)
=e^{i\frac{\eta^3z}{2}\left(x-\frac{\eta^3z^2}{6}\right)} \mathrm{Ai}\left[\eta\left(x-\frac{\eta^3z^2}{4}\right)\right].
\end{equation}
Hence, in a manner analogous to the previous case, we can derive an initial expression for $e^{\Delta(z)\hat{p}_x}\mathrm{Ai}(\eta x)$ by substituting $z \rightarrow 2i\Delta(z)$, such that
\begin{equation}
    \begin{split}
e^{\Delta(z)\hat{p}_x^2} \mathrm{Ai}(\eta x)
&=e^{-{\eta^3\Delta(z)}\left[x+\frac{2\eta^3\Delta^2(z)}{3}\right]}\\&\times 
\mathrm{Ai}\left[\eta\left(x+{\eta^3\Delta^2(z)}\right)\right].
    \end{split}
\end{equation}
Consequently, from Eq.~(\ref{eq:0080}), and using the commutation relations of the operators $\hat{p}_q^2$, $q^2$, and $q\hat{p}_q+\hat{p}_qq$ (with $q=x,y$), which are closed under commutation, we obtain that
\begin{equation}\label{Airy}
\begin{split}
E\left(x,y,z\right)&=\frac{e^{-\left(\frac{g_xx^2}{w_x(z)}+\frac{g_yy^2}{w_y(z)}\right)}}{\sqrt{w_x(z)w_y(z)}}\\&\times
e^{ -{\eta^3\Delta(z)}\left[\frac{x}{w_x(z)}+i\frac{y}{w_y(z)}+\frac{2\eta^3\Delta^2(z)}{3}\right]}
\\&\times 
\mathrm{Ai}\left[\eta\left(\frac{x}{w_x(z)}+i\frac{y}{w_y(z)}+{\eta^3\Delta^2(z)}\right)\right].
\end{split}
\end{equation}
Fig.~\ref{Fig_2} illustrates the propagation of an Airy function, $\mathcal{E}(x+iy)=\mathrm{Ai}[\eta(x+iy)]$. Similar to the case of the Bessel function, we present three transverse observation planes showing the intensity distribution of the propagated field, both analytically and experimentally through the synthesis of the optical field in question. Consequently, from Eq.~(\ref{Airy}), the first column illustrates the optical beam intensity at different distances: $(\text{a}_1)$ at $z=0\,\text{m}$; $(\text{b}_1)$ at $z=0.20\,\text{m}$; and $(\text{c}_1)$ at $z=0.90\,\text{m}$. On the other hand, the second column in Fig.~\ref{Fig_2} illustrates the intensity evolution at the same propagation distances and under identical initial conditions, utilizing the experimentally synthesized optical field detailed in Fig.~\ref{Fig_0}.
This comparative analysis highlights the asymmetry introduced by the Gaussian modulation. As observed, similar to the previous case, the ratio $g_y/g_x = 5.0$ not only causes a rotation along the propagation direction but also induces a pronounced asymmetric distortion, which is clearly attributed to the asymmetric Gaussian modulation. The close agreement between theoretical predictions and experimental results not only validates the theoretical model but also underscores its practical significance, confirming the efficacy of the optical field synthesis technique~\cite{Arrizon07}.

\textit{Conclusions.} We have introduced an innovative approach to solving the paraxial equation by applying operator techniques from quantum mechanics, specifically quantum optics. This approach leverages the fact that $\nabla_{\perp}^2 \mathcal{E}(x+iy)=0$. The asymmetric nature of Gaussian modulation poses challenges making the direct application of such techniques less straightforward compared to previous methods like those in reference \cite{Moya_2024}. However, by reducing the problem from two-dimensional to one-dimensional propagation, we have derived a useful integral representation for an asymmetrically modulated entire function $\mathcal{E}(x+iy)$. This reduction not only simplifies the problem but also enhances our ability to analyze complex optical fields. Our investigation, that includes both analytical and experimental studies of the Bessel and Airy functions, confirms the validity and utility of this approach, suggesting its potential for further research and practical applications in optical field synthesis, quantitative phase imaging, and other related areas~\cite{Ashkin_1987,Oh_2023}.

%
    
\end{document}